\documentclass{article}
\usepackage{amsfonts,amssymb,amsmath,amsthm,graphicx}
\usepackage[english]{babel}
\let\phi=\varphi

\let\theta=\vartheta

\newcommand{\bC}{{\mathbb C}}

\newcommand{\bZ}{{\mathbb Z}}

\newcommand{\ket}{\rangle}

\begin{document}
\title{\Large {\bf A Sequence of Qubit-Qudit Pauli Groups as a Nested Structure of Doilies}}

\author{Metod Saniga$^{1}$ and Michel Planat$^{2}$\\
\normalsize $^{1}$Astronomical Institute, Slovak Academy of Sciences\\
\normalsize  SK-05960 Tatransk\' a Lomnica, Slovak Republic\\
\normalsize  (msaniga@astro.sk)
\\
\normalsize  $^{2}$Institut FEMTO-ST, CNRS, 32 Avenue de l'Observatoire\\
\normalsize  F-25044 Besan\c con Cedex, France\\
\normalsize  (michel.planat@femto-st.fr)}

\date{\small (12 April 2011)}

\maketitle

\begin{abstract}\noindent
Following the spirit of a recent work of one of the authors (J. Phys. A: Math. Theor. 44 (2011) 045301), the essential structure of the generalized Pauli group of a qubit-qu$d$it, where $d = 2^{k}$ and an integer $k \geq 2$, is recast in the language of a finite geometry. A point of such geometry is represented by the maximum set of mutually commuting elements of the group and two distinct points are regarded as collinear if the corresponding sets have exactly $2^{k} - 1$ elements in common. The geometry comprises $2^{k} - 1$ copies of the generalized quadrangle of order two (``the doily") that form $2^{k-1} - 1$ pencils arranged into a remarkable nested configuration. This nested structure reflects the fact that maximum sets of mutually commuting elements are of two different kinds (ordinary and exceptional) and exhibits an intriguing alternating pattern: the subgeometry of the exceptional points of the $(k+2)$-case is found to be isomorphic to the full geometry of the $k$-case. It should be stressed, however, that these generic properties of the qubit-qudit geometry were inferred from purely computer-handled cases of $k = 2, 3, 4$ and $5$ only and, therefore, their rigorous, computer-free proof for $k \geq 6$ still remains a mathematical challenge. \\


~~~~~~~~~~~-- Nested Structure

\end{abstract}

\vspace*{.5cm}
\section{Introduction}
Although finite geometries/point-line incidence structures behind the generalized Pauli groups of {\it single} qudits have essentially all been discovered and thoroughly explored \cite{hs07}--\cite{thas}, only little is still known about those underlying the groups of  {\it multi}-qudits and/or tensor products of qudits of {\it different} ranks. A number of particular cases have been analyzed by computer \cite{pb}--\cite{pl2}, and certain ``easy-to-handle" configurations have also been studied and completely described in a computer-free fashion \cite{thas},\cite{ps08}, \cite{sp07}--\cite{vrlev}.  Based on the recent work of one of us \cite{pl2}, we shall address below a particular case of the tensor product of a qubit and a single qudit, where  $d = 2^{k}$ and an integer $k \geq 2$. Analyzing the first four cases in the sequence, viz. $k=2, 3, 4$ and $5$, by computer was already sufficient for us to infer the generic pattern of geometries behind the corresponding Pauli groups for any $k \geq 2$. This sequence of finite geometries is characterized by a remarkable nesting of fundamental building blocks that are nothing but generalized quadrangles of order two, GQ(2,\,2)s. This seems to be the most crucial finding given the facts that this generalized quadrangle 1) is the geometry behind the generalized Pauli group of {\it two-qubits} \cite{ps08,spp08} and 2) also plays, together with other prominent point-line incidence structures, an essential role in finite geometrical aspects of the still mysterious black-hole/qubit correspondence \cite{lsvp,lsv}.

\section{Qudits, Generalized Pauli Groups and the Doily}
Given an integer $d>1$ and $\bZ_d:=\{0,1,\ldots,d-1\}$, where addition and
multiplication of elements from $\bZ_d$ is understood modulo $d$,
we consider the $d$-dimensional complex Hilbert space $\bC^d$ and denote by
\begin{equation*}
   \{\, |s\ket : s\in\bZ_d \}
\end{equation*}
a computational basis of $\bC^d$. A qudit in $\bC^{d}$ is a vector
\begin{equation*}
| \psi \ket = \sum^{d-1}_{s=0} \alpha_s | s \ket,~{\rm where}~\alpha_s \in \bC~{\rm and}~\sum^{d-1}_{s=0} |\alpha_s|^{2} = 1.
\end{equation*}
Taking $\omega_{(d)}$ to be a fixed
primitive $d$-th root of unity (e.\,g.,\ $\omega_{(d)}=\exp(2\pi i/d)$), we define
unitary $X_{(d)}$ (``shift") and $Z_{(d)}$ (``clock") operators on $\bC^d$ via $X_{(d)}|s\ket =
|s+1\ket$ and $Z_{(d)}|s\ket = \omega_{(d)}^s |s\ket$ for all $s\in\bZ_d$.
In the computational basis
\begin{equation}\label{eq:xz}
  X_{(d)} =  \begin{pmatrix}
    0      & 0     &\ldots &0      & 1\\
    1      & 0     &\ldots &0      & 0\\
    0      & 1     &\ldots &0      & 0\\
    \vdots &\vdots &\ddots &\vdots &\vdots\\
    0      & 0     &\ldots &1      & 0
    \end{pmatrix}
    \mbox{~~~~and~~~~}
  Z_{(d)} =  \begin{pmatrix}
    1      & 0     &0        &\ldots&  0\\
    0      &\omega &0        &\ldots&  0\\
    0      & 0     &\omega^2 &\ldots& 0\\
    \vdots &\vdots &\vdots   &\ddots &\vdots\\
    0  & 0 & 0     &\ldots   &\omega^{d-1}
    \end{pmatrix}.
\end{equation}
The (generalized) \emph{Pauli group} generated by $X_{(d)}$ and $Z_{(d)}$ will be
denoted as $G_{(d)}$.
For all $s\in\bZ_d$ we have $X_{(d)}Z_{(d)}|s\ket = \omega_{(d)}^s|s+1\ket$ and $Z_{(d)}X_{(d)}|s\ket =
\omega_{(d)}^{s+1}|s+1\ket$. This gives the basic relation
\begin{equation*}\label{eq:xzzx}
    \omega_{(d)} X_{(d)}Z_{(d)} = Z_{(d)}X_{(d)}
\end{equation*}
which implies that each element of $G_{(d)}$ can be
written in the unique \emph{normal form}
\begin{equation}\label{eq:normalform}
    \omega_{(d)}^a X_{(d)}^b Z_{(d)}^c \mbox{~~for some integers~~}a,b,c\in \bZ_d.
\end{equation}
The uniqueness of this normal form implies that $G_{(d)}$ is a group of order $d^3$.
From (\ref{eq:xzzx}) it is also readily seen that
\begin{equation*}
    (\omega_{(d)}^a X_{(d)}^b Z_{(d)}^c) (\omega_{(d)}^{a'} X_{(d)}^{b'} Z_{(d)}^{c'}) = \omega_{(d)}^{b'c + a+a'}
X_{(d)}^{b+b'} Z_{(d)}^{c+c'}.
\end{equation*}
which implies that \emph{commutator\/} of two operators $W$ and $W'$,
\begin{equation*}\label{eq:defcommutator}
    [W,W'] := W W'W^{-1}{W'} ^{-1},
\end{equation*}
acquires in our case the form
\begin{equation}\label{eq:commutator}
    [\omega_{(d)}^a X_{(d)}^b Z_{(d)}^c, \omega_{(d)}^{a'} X_{(d)}^{b'}Z_{(d)}^{c'}] = \omega_{(d)}^{cb'-c'b}I_{(d)}.
\end{equation}
Let us recall that two operators commute if, and only if, their commutator (taken in
any order) is equal to $I_{(d)}$ (the identity matrix); hence, $G_{(d)}$ is a {\emph{non}-commutative group.  We also mention in passing that
there are two important normal subgroups of $G_{(d)}$:
its \emph{centre\/} $Z(G_{(d)})$ and its \emph{commutator subgroup} $G_{(d)}'$, the two being identical
\begin{equation*}\label{eq:G'}
    G_{(d)}' = Z(G_{(d)}) = \{\omega_{(d)}^a I_{(d)}: a\in\bZ_d \}.
\end{equation*}
It is important to observe that $a$ and $a'$ do not occur on the right-hand side of eq.\,(\ref{eq:commutator}). So, in order to study the commutation relations between the elements of $G_{(d)}$ one can disregard the complex phase factors $\omega_{(d)}^{a}$ and work \emph{solely} with $d^2$ matrix products $X_{(d)}^bZ_{(d)}^c$.

The final notion we shall need in the sequel is that of a \emph{finite generalized quadrangle} of order $(s, t)$, usually denoted GQ($s, t$). This is an incidence structure $S = (P, B, {\rm I})$,
where $P$ and $B$ are disjoint (non-empty) sets of objects, called respectively points and lines, and where I is a symmetric point-line
incidence relation satisfying the following axioms \cite{pay-thas}: (i) each point is incident with $1 + t$ lines ($t \geq 1$) and two
distinct points are incident with at most one line; (ii) each line is incident with $1 + s$ points ($s \geq 1$) and two distinct lines
are incident with at most one point;  and (iii) if $x$ is a point and $L$ is a line not incident with $x$, then there exists a unique
pair $(y, M) \in  P \times B$ for which $x {\rm I} M {\rm I} y {\rm I} L$; from these axioms it readily follows that $|P| = (s+1)(st+1)$
and $|B| = (t+1)(st+1)$.  If $s = t$, $S$ is said to have order $s$. A
generalized quadrangle with both $s > 1$ and $t > 1$ is called thick. The smallest thick generalized quadrangle is obviously the (unique) GQ(2,\,2), often dubbed the ``doily." This quadrangle is endowed with
15 points/lines, with each line containing three points and, dually, each point being on three lines. Moreover, it is a self-dual object,
i.\,e., isomorphic to its dual. In the older literature this geometry is also known as the Cremona-Richmond configuration (see, e.\,g., \cite{cox}) and essential features of its structure are depicted in Figure 1.
\begin{figure}[pth!]
\centerline{\includegraphics[width=6.cm,clip=]{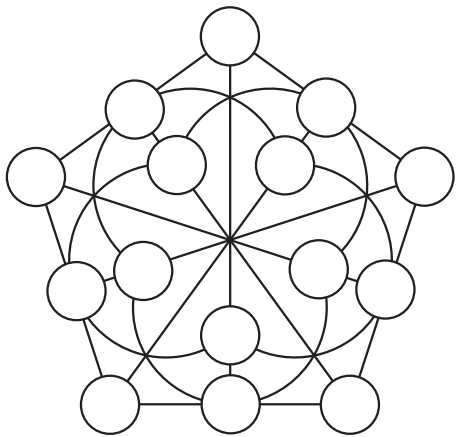}}
\vspace*{.2cm}
\caption{A picture of the generalized quadrangle of order two, the doily. The circles stand for its points, while its lines are given by triples of points lying on the same segments (10) and/or arcs of circles (5). (Note that this point-line incidence geometry does not contain any triangles.)}
\end{figure}

\section{The Qubit-Qu$2^k$it Pauli Group and Its Geometry}
As already mentioned, we shall focus on the generalized Pauli group associated with the Hilbert space of the form $\bC^2 \otimes \bC^{2^k}$, where an integer $k \geq 2$.
\par
We shall first handle the $k=2$ case. Disregarding the phase factors, from eqs.\,(\ref{eq:xz}) and (\ref{eq:normalform}) one sees that the elements of the group associated with $\bC^2$ (qubit) are
\begin{equation*}
\{I_{(2)}, X_{(2)}, Z_{(2)}, X_{(2)}Z_{(2)}\},
\end{equation*}
and those for $\bC^{4}$ read
\begin{eqnarray*}
&&\{I_{(4)}, X_{(4)}, X_{(4)}^2, X_{(4)}^3, Z_{(4)}, X_{(4)}Z_{(4)}, X_{(4)}^2Z_{(4)}, X_{(4)}^3Z_{(4)}, Z_{(4)}^2, \nonumber \\
  &&X_{(4)}Z_{(4)}^2, X_{(4)}^2Z_{(4)}^2, X_{(4)}^3Z_{(4)}^2, Z_{(4)}^3, X_{(4)}Z_{(4)}^3, X_{(4)}^2Z_{(4)}^3, X_{(4)}^3Z_{(4)}^3\}.
\end{eqnarray*}
From the last two equations it follows that the generalized Pauli group of $\bC^2 \otimes \bC^{4}$ will feature the following elements (again disregarding the phase factors)
\begin{eqnarray*}
&&\{I_{(2)} \otimes  I_{(4)}, I_{(2)} \otimes  X_{(4)}, I_{(2)} \otimes  X_{(4)}^2,\ldots, I_{(2)} \otimes X_{(4)}^2Z_{(4)}^3, I_{(2)} \otimes X_{(4)}^3Z_{(4)}^3, \nonumber \\
  && X_{(2)} \otimes  I_{(4)}, X_{(2)} \otimes  X_{(4)}, X_{(2)} \otimes  X_{(4)}^2,\ldots, X_{(2)} \otimes X_{(4)}^2Z_{(4)}^3, X_{(2)} \otimes X_{(4)}^3Z_{(4)}^3, \nonumber \\
 && Z_{(2)} \otimes  I_{(4)}, Z_{(2)} \otimes  X_{(4)}, Z_{(2)} \otimes  X_{(4)}^2,\ldots, Z_{(2)} \otimes X_{(4)}^2Z_{(4)}^3, Z_{(2)} \otimes X_{(4)}^3Z_{(4)}^3, \nonumber \\
 && X_{(2)}Z_{(2)} \otimes  I_{(4)}, X_{(2)}Z_{(2)} \otimes  X_{(4)}, X_{(2)}Z_{(2)} \otimes  X_{(4)}^2,\ldots, X_{(2)}Z_{(2)} \otimes X_{(4)}^3Z_{(4)}^3\}.
\end{eqnarray*}
Since $I_{(2)} \otimes  I_{(4)}$ commutes with every other element, we consider only the remaining 63 elements/operators and, for convenience, number them from 1 to 63 in the order as they appear in the last equation.

\par
To find the geometry behind this Pauli group, we first look for maximum sets of mutually commuting elements. We find that each such set features seven elements and their total number is 39.
Three of them, namely
\begin{eqnarray*}\label{eq:exceptional}
&& a \equiv \{2, 8, 10, 16, 18, 24, 26\},\nonumber \\
&& b \equiv \{2, 8, 10, 32, 34, 40, 42\},\\
&& c \equiv \{2, 8, 10, 48, 50, 56, 58\}, \nonumber
\end{eqnarray*}
are special (and henceforth referred to as exceptional) in the sense that each of them shares with any other set at least one element. The remaining 36 sets split into three distinct, equally-sized families, in particular family I
\begin{eqnarray*}
1^{\star} &\equiv& \{1, 2, 3, 16, 17, 18, 19\}, \nonumber \\
2^{\star} &\equiv& \{1, 2, 3, 32, 33, 34, 35\}, \nonumber \\
3^{\star} &\equiv& \{1, 2, 3, 48, 49, 50, 51\}, \nonumber \\
4^{\star} &\equiv& \{2, 9, 11, 16, 18, 25, 27\}, \nonumber \\
5^{\star} &\equiv& \{2, 9, 11, 32, 34, 41, 43\}, \nonumber \\
6^{\star} &\equiv& \{2, 9, 11, 48, 50, 57, 59\},\\
7^{\star} &\equiv& \{2, 24, 26, 33, 35, 57, 59\}, \nonumber \\
8^{\star} &\equiv& \{2, 24, 26, 41, 43, 49, 51\}, \nonumber \\
9^{\star} &\equiv& \{2, 25, 27, 33, 35, 56, 58\}, \nonumber \\
10^{\star} &\equiv& \{2, 25, 27, 40, 42, 49, 51\}, \nonumber \\
11^{\star} &\equiv& \{2, 17, 19, 40, 42, 57, 59\}, \nonumber \\
12^{\star} &\equiv& \{2, 17, 19, 41, 43, 56, 58\}, \nonumber
\end{eqnarray*}
family II
\begin{eqnarray*}
1^{\bullet} &\equiv& \{4, 8, 12, 16, 20, 24, 28\}, \nonumber \\
2^{\bullet} &\equiv& \{4, 8, 12, 32, 36, 40, 44\}, \nonumber \\
3^{\bullet} &\equiv& \{4, 8, 12, 48, 52, 56, 60\}, \nonumber \\
4^{\bullet} &\equiv& \{6, 8, 14, 16, 22, 24, 30\}, \nonumber \\
5^{\bullet} &\equiv& \{6, 8, 14, 32, 38, 40, 46\}, \nonumber \\
6^{\bullet} &\equiv& \{6, 8, 14, 48, 54, 56, 62\}, \\
7^{\bullet} &\equiv& \{8, 20, 28, 34, 42, 54, 62\}, \nonumber \\
8^{\bullet} &\equiv& \{8, 20, 28, 38, 46, 50, 58\}, \nonumber \\
9^{\bullet} &\equiv& \{8, 22, 30, 34, 42, 52, 60\}, \nonumber \\
10^{\bullet} &\equiv& \{8, 22, 30, 36, 44, 50, 58\}, \nonumber \\
11^{\bullet} &\equiv& \{8, 18, 26, 36, 44, 54, 62\}, \nonumber \\
12^{\bullet} &\equiv& \{8, 18, 26, 38, 46, 52, 60\}, \nonumber
\end{eqnarray*}
and family III
\begin{eqnarray*}
1' &\equiv& \{5, 10, 15, 16, 21, 26, 31\}, \nonumber \\
2' &\equiv& \{5, 10, 15, 32, 37, 42, 47\}, \nonumber \\
3' &\equiv& \{5, 10, 15, 48, 53, 58, 63\}, \nonumber \\
4' &\equiv& \{7, 10, 13, 16, 23, 26, 29\}, \nonumber \\
5' &\equiv& \{7, 10, 13, 32, 39, 42, 45\}, \nonumber \\
6' &\equiv& \{7, 10, 13, 48, 55, 58, 61\}, \\
7' &\equiv& \{10, 21, 31, 34, 40, 55, 61\}, \nonumber \\
8' &\equiv& \{10, 21, 31, 39, 45, 50, 56\}, \nonumber \\
9' &\equiv& \{10, 23, 29, 34, 40, 53, 63\}, \nonumber \\
10' &\equiv& \{10, 23, 29, 37, 47, 50, 56\}, \nonumber \\
11' &\equiv& \{10, 18, 24, 37, 47, 55, 61\}, \nonumber \\
12' &\equiv& \{10, 18, 24, 39, 45, 53, 63\}. \nonumber
\end{eqnarray*}
It is easy to verify that this split into three above-given families is unique as the twelve sets in each family have a single  element in common ($2 = I_{(2)} \otimes X_{(4)}^2$, $8 = I_{(2)} \otimes Z_{(4)}^2$ and $10 = I_{(2)} \otimes X_{(4)}^2 Z_{(4)}^2$, respectively).

\par
Next, we define the point-line incidence geometry whose points are these 39 maximum sets of mutually commuting elements and where two points are collinear if the corresponding sets have exactly \emph{three} ($2^2 - 1$) elements in common. The three exceptional sets  then form a distinguished line in this geometry and each family generates a copy of the doily from which one line is deleted; this line being nothing but the distinguished line! Hence, this point-line incidence structure comprises three doilies on a common line, a pencil of doilies --- as illustrated in Figure 2.
\begin{figure}[pth!]
\centerline{\includegraphics[width=15cm,clip=]{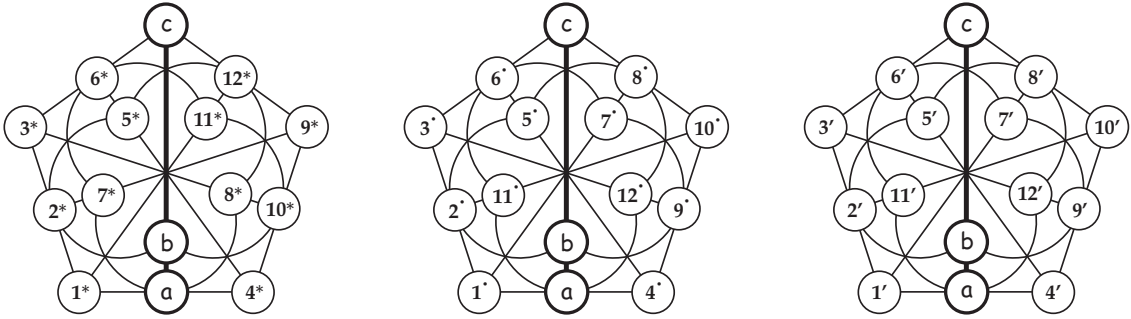}}
\vspace*{.2cm} \caption{A diagrammatical illustration of the finite geometry behind the qubit-ququartit Pauli group: a set of three generalized quadrangles of order two sharing a line.}
\end{figure}

\par
In a completely analogous way, we shall analyze the $k = 3$ case. Here, the ``relevant" elements of the Pauli group related to the Hilbert space $\bC^{8}$ are (after dropping the subscript ``8" for better readability of the formulas)
\begin{eqnarray*}
&&\{I, X, X^2, X^3, X^4, X^5, X^6, X^7, Z, ZX, ZX^2, ZX^3, ZX^4, ZX^5, ZX^6, ZX^7, Z^2, Z^2X, \nonumber \\
&& Z^2X^2, Z^2X^3, Z^2X^4, Z^2X^5, Z^2X^6, Z^2X^7, Z^3, Z^3X, Z^3X^2, Z^3X^3, Z^3X^4, Z^3X^5, \nonumber \\
&& Z^3X^6, Z^3X^7, Z^4, Z^4X, Z^4X^2, Z^4X^3, Z^4X^4, Z^4X^5, Z^4X^6, Z^4X^7, Z^5, Z^5X, Z^5X^2, \nonumber \\
&& Z^5X^3, Z^5X^4, Z^5X^5, Z^5X^6, Z^5X^7, Z^6, Z^6X, Z^6X^2, Z^6X^3, Z^6X^4, Z^6X^5, Z^6X^6, Z^6X^7, \nonumber \\
&& Z^7, Z^7X, Z^7X^2, Z^7X^3, Z^7X^4, Z^7X^5, Z^7X^6, Z^7X^7\},
\end{eqnarray*}
which yield the following 255 (disregarding the identity) distinct matrices of the group associated with $\bC^{2} \otimes \bC^{8}$
\begin{eqnarray*}\label{eq:255}
&&\{I_{(2)} \otimes X, I_{(2)} \otimes X^2,\dots, I_{(2)} \otimes  Z^7X^7, X_{(2)} \otimes I, X_{(2)} \otimes X,\dots, X_{(2)}\otimes  Z^7X^7, \nonumber \\
&&Z_{(2)}\otimes I, Z_{(2)}\otimes X,\dots, Z_{(2)} \otimes  Z^7X^7, X_{(2)}Z_{(2)}\otimes I,\dots, X_{(2)}Z_{(2)} \otimes  Z^7X^7\}.
\end{eqnarray*}
Again, to facilitate our reasoning, we number these elements from 1 to 255 consecutively as they are listed in the last expression. Using computer, we find out that they form 87 maximum sets of mutually commuting guys, each of cardinality 15. Fifteen of them are exceptional and they explicitly read
\begin{eqnarray*}
a &\equiv& \{2, 4, 6, 32, 34, 36, 38, 64, 66, 68, 70, 96, 98, 100, 102\}, \nonumber \\
b &\equiv& \{2, 4, 6, 32, 34, 36, 38, 128, 130, 132, 134, 160, 162, 164, 166\}, \nonumber \\
c &\equiv& \{2, 4, 6, 32, 34, 36, 38, 192, 194, 196, 198, 224, 226, 228, 230\}, \nonumber \\
d &\equiv& \{4, 16, 20, 32, 36, 48, 52, 64, 68, 80, 84, 96, 100, 112, 116\}, \nonumber \\
e &\equiv& \{ 4, 16, 20, 32, 36, 48, 52, 128, 132, 144, 148, 160, 164, 176, 180\}, \nonumber \\
f &\equiv& \{ 4, 16, 20, 32, 36, 48, 52, 192, 196, 208, 212, 224, 228, 240, 244\}, \nonumber \\
g &\equiv& \{ 4, 18, 22, 32, 36, 50, 54, 64, 68, 82, 86, 96, 100, 114, 118\}, \nonumber \\
h &\equiv& \{ 4, 18, 22, 32, 36, 50, 54, 128, 132, 146, 150, 160, 164, 178, 182\}, \\
i &\equiv& \{ 4, 18, 22, 32, 36, 50, 54, 192, 196, 210, 214, 224, 228, 242, 246\}, \nonumber \\
j &\equiv& \{ 4, 32, 36, 80, 84, 112, 116, 130, 134, 162, 166, 210, 214, 242, 246\}, \nonumber \\
k &\equiv& \{ 4, 32, 36, 80, 84, 112, 116, 146, 150, 178, 182, 194, 198, 226, 230\}, \nonumber \\
l &\equiv& \{ 4, 32, 36, 82, 86, 114, 118, 130, 134, 162, 166, 208, 212, 240, 244\}, \nonumber \\
m &\equiv& \{ 4, 32, 36, 82, 86, 114, 118, 144, 148, 176, 180, 194, 198, 226, 230\}, \nonumber \\
n &\equiv& \{ 4, 32, 36, 66, 70, 98, 102, 144, 148, 176, 180, 210, 214, 242, 246\}, \nonumber \\
o &\equiv& \{ 4, 32, 36, 66, 70, 98, 102, 146, 150, 178, 182, 208, 212, 240, 244\}. \nonumber
\end{eqnarray*}
The remaining 72 ``ordinary" sets split, as expected, into six sets of cardinality twelve each. These six sets are found to form three {\it pairs}, namely pair $A$
\begin{eqnarray*}
1 &\equiv& \{1, 2, 3, 4, 5, 6, 7, 64, 65, 66, 67, 68, 69, 70, 71\},      \nonumber \\
2 &\equiv& \{1, 2, 3, 4, 5, 6, 7, 128, 129, 130, 131, 132, 133, 134, 135\}, \nonumber \\
3 &\equiv& \{1, 2, 3, 4, 5, 6, 7, 192, 193, 194, 195, 196, 197, 198, 199\}, \nonumber \\
4 &\equiv& \{2, 4, 6, 33, 35, 37, 39, 64, 66, 68, 70, 97, 99, 101, 103\}, \nonumber \\
5 &\equiv& \{2, 4, 6, 33, 35, 37, 39, 128, 130, 132, 134, 161, 163, 165, 167\}, \nonumber \\
6 &\equiv& \{2, 4, 6, 33, 35, 37, 39, 192, 194, 196, 198, 225, 227, 229, 231\}, \nonumber \\
7 &\equiv& \{2, 4, 6, 96, 98, 100, 102, 129, 131, 133, 135, 225, 227, 229, 231\}, \nonumber \\
8 &\equiv& \{2, 4, 6, 96, 98, 100, 102, 161, 163, 165, 167, 193, 195, 197, 199\}, \nonumber \\
9 &\equiv& \{2, 4, 6, 97, 99, 101, 103, 129, 131, 133, 135, 224, 226, 228, 230\}, \nonumber \\
10 &\equiv& \{2, 4, 6, 97, 99, 101, 103, 160, 162, 164, 166, 193, 195, 197, 199\}, \nonumber \\
11 &\equiv& \{2, 4, 6, 65, 67, 69, 71, 160, 162, 164, 166, 225, 227, 229, 231\}, \nonumber \\
12 &\equiv& \{2, 4, 6, 65, 67, 69, 71, 161, 163, 165, 167, 224, 226, 228, 230\}, \nonumber
\end{eqnarray*}
\begin{eqnarray*}
1' &\equiv&     \{4, 17, 21, 34, 38, 51, 55, 64, 68, 81, 85, 98, 102, 115, 119\},         \nonumber \\
2' &\equiv&    \{4, 17, 21, 34, 38, 51, 55, 128, 132, 145, 149, 162, 166, 179, 183\}, \nonumber \\
3' &\equiv&    \{4, 17, 21, 34, 38, 51, 55, 192, 196, 209, 213, 226, 230, 243, 247\}, \nonumber \\
4' &\equiv&    \{4, 19, 23, 34, 38, 49, 53, 64, 68, 83, 87, 98, 102, 113, 117\}, \nonumber \\
5' &\equiv&    \{4, 19, 23, 34, 38, 49, 53, 128, 132, 147, 151, 162, 166, 177, 181\}, \nonumber \\
6' &\equiv&    \{4, 19, 23, 34, 38, 49, 53, 192, 196, 211, 215, 226, 230, 241, 245\}, \nonumber \\
7' &\equiv&   \{4, 34, 38, 81, 85, 115, 119, 130, 134, 160, 164, 211, 215, 241, 245\}, \nonumber \\
8' &\equiv&    \{4, 34, 38, 81, 85, 115, 119, 147, 151, 177, 181, 194, 198, 224, 228\}, \nonumber \\
9' &\equiv&    \{4, 34, 38, 83, 87, 113, 117, 130, 134, 160, 164, 209, 213, 243, 247\}, \nonumber \\
10' &\equiv&    \{4, 34, 38, 83, 87, 113, 117, 145, 149, 179, 183, 194, 198, 224, 228\}, \nonumber \\
11' &\equiv&    \{4, 34, 38, 66, 70, 96, 100, 145, 149, 179, 183, 211, 215, 241, 245\}, \nonumber \\
12' &\equiv&   \{4, 34, 38, 66, 70, 96, 100, 147, 151, 177, 181, 209, 213, 243, 247\}, \nonumber
\end{eqnarray*}
pair $B$
\begin{eqnarray*}
1^{\bullet} &\equiv& \{8, 16, 24, 32, 40, 48, 56, 64, 72, 80, 88, 96, 104, 112, 120\},          \nonumber \\
2^{\bullet} &\equiv& \{8, 16, 24, 32, 40, 48, 56, 128, 136, 144, 152, 160, 168, 176, 184\}, \nonumber \\
3^{\bullet} &\equiv& \{8, 16, 24, 32, 40, 48, 56, 192, 200, 208, 216, 224, 232, 240, 248\}, \nonumber \\
4^{\bullet} &\equiv& \{12, 16, 28, 32, 44, 48, 60, 128, 140, 144, 156, 160, 172, 176, 188\}, \nonumber \\
5^{\bullet} &\equiv& \{12, 16, 28, 32, 44, 48, 60, 192, 204, 208, 220, 224, 236, 240, 252\}, \nonumber \\
6^{\bullet} &\equiv& \{12, 16, 28, 32, 44, 48, 60, 64, 76, 80, 92, 96, 108, 112, 124\},     \nonumber \\
7^{\bullet} &\equiv& \{16, 32, 48, 72, 88, 104, 120, 132, 148, 164, 180, 204, 220, 236, 252\}, \nonumber \\
8^{\bullet} &\equiv& \{16, 32, 48, 72, 88, 104, 120, 140, 156, 172, 188, 196, 212, 228, 244\}, \nonumber \\
9^{\bullet} &\equiv& \{16, 32, 48, 76, 92, 108, 124, 132, 148, 164, 180, 200, 216, 232, 248\}, \nonumber \\
10^{\bullet} &\equiv& \{16, 32, 48, 76, 92, 108, 124, 136, 152, 168, 184, 196, 212, 228, 244\}, \nonumber \\
11^{\bullet} &\equiv& \{16, 32, 48, 68, 84, 100, 116, 140, 156, 172, 188, 200, 216, 232, 248\},  \nonumber \\
12^{\bullet} &\equiv& \{16, 32, 48, 68, 84, 100, 116, 136, 152, 168, 184, 204, 220, 236, 252\}, \nonumber
\end{eqnarray*}
\begin{eqnarray*}
\overline{1} &\equiv&  \{10, 20, 30, 32, 42, 52, 62, 128, 138, 148, 158, 160, 170, 180, 190\},    \nonumber \\
\overline{2} &\equiv& \{10, 20, 30, 32, 42, 52, 62, 192, 202, 212, 222, 224, 234, 244, 254\},     \nonumber \\
\overline{3} &\equiv& \{14, 20, 26, 32, 46, 52, 58, 64, 78, 84, 90, 96, 110, 116, 122\}, \nonumber \\
\overline{4} &\equiv& \{14, 20, 26, 32, 46, 52, 58, 128, 142, 148, 154, 160, 174, 180, 186\}, \nonumber \\
\overline{5} &\equiv& \{14, 20, 26, 32, 46, 52, 58, 192, 206, 212, 218, 224, 238, 244, 250\},    \nonumber \\
\overline{6} &\equiv& \{10, 20, 30, 32, 42, 52, 62, 64, 74, 84, 94, 96, 106, 116, 126\}, \nonumber \\
\overline{7} &\equiv& \{20, 32, 52, 74, 94, 106, 126, 132, 144, 164, 176, 206, 218, 238, 250\}, \nonumber \\
\overline{8} &\equiv& \{20, 32, 52, 74, 94, 106, 126, 142, 154, 174, 186, 196, 208, 228, 240\},    \nonumber \\
\overline{9} &\equiv& \{20, 32, 52, 78, 90, 110, 122, 132, 144, 164, 176, 202, 222, 234, 254\}, \nonumber \\
\overline{10} &\equiv& \{20, 32, 52, 78, 90, 110, 122, 138, 158, 170, 190, 196, 208, 228, 240\},  \nonumber \\
\overline{11} &\equiv& \{20, 32, 52, 68, 80, 100, 112, 138, 158, 170, 190, 206, 218, 238, 250\},  \nonumber \\
\overline{12} &\equiv& \{20, 32, 52, 68, 80, 100, 112, 142, 154, 174, 186, 202, 222, 234, 254\}, \nonumber
\end{eqnarray*}
and pair $C$
\begin{eqnarray*}
1^{\star} &\equiv& \{9, 18, 27, 36, 45, 54, 63, 64, 73, 82, 91, 100, 109, 118, 127\},          \nonumber \\
2^{\star} &\equiv& \{9, 18, 27, 36, 45, 54, 63, 128, 137, 146, 155, 164, 173, 182, 191\}, \nonumber \\
3^{\star} &\equiv& \{9, 18, 27, 36, 45, 54, 63, 192, 201, 210, 219, 228, 237, 246, 255\}, \nonumber \\
4^{\star} &\equiv& \{18, 36, 54, 73, 91, 109, 127, 132, 150, 160, 178, 205, 223, 233, 251\}, \nonumber \\
5^{\star} &\equiv& \{18, 36, 54, 73, 91, 109, 127, 141, 159, 169, 187, 196, 214, 224, 242\}, \nonumber \\
6^{\star} &\equiv& \{18, 36, 54, 77, 95, 105, 123, 132, 150, 160, 178, 201, 219, 237, 255\}, \nonumber \\
7^{\star} &\equiv& \{18, 36, 54, 77, 95, 105, 123, 137, 155, 173, 191, 196, 214, 224, 242\}, \nonumber \\
8^{\star} &\equiv& \{18, 36, 54, 68, 86, 96, 114, 137, 155, 173, 191, 205, 223, 233, 251\}, \nonumber \\
9^{\star} &\equiv& \{13, 18, 31, 36, 41, 54, 59, 64, 77, 82, 95, 100, 105, 118, 123\}, \nonumber \\
10^{\star} &\equiv& \{13, 18, 31, 36, 41, 54, 59, 128, 141, 146, 159, 164, 169, 182, 187\}, \nonumber \\
11^{\star} &\equiv& \{13, 18, 31, 36, 41, 54, 59, 192, 205, 210, 223, 228, 233, 246, 251\}, \nonumber \\
12^{\star} &\equiv& \{18, 36, 54, 68, 86, 96, 114, 141, 159, 169, 187, 201, 219, 237, 255\}, \nonumber
\end{eqnarray*}
\begin{eqnarray*}
\widetilde{1} &\equiv&  \{11, 22, 25, 36, 47, 50, 61, 64, 75, 86, 89, 100, 111, 114, 125\},       \nonumber \\
\widetilde{2} &\equiv& \{11, 22, 25, 36, 47, 50, 61, 128, 139, 150, 153, 164, 175, 178, 189\}, \nonumber \\
\widetilde{3} &\equiv& \{11, 22, 25, 36, 47, 50, 61, 192, 203, 214, 217, 228, 239, 242, 253\},    \nonumber \\
\widetilde{4} &\equiv& \{22, 36, 50, 75, 89, 111, 125, 132, 146, 160, 182, 207, 221, 235, 249\}, \nonumber \\
\widetilde{5} &\equiv& \{22, 36, 50, 75, 89, 111, 125, 143, 157, 171, 185, 196, 210, 224, 246\},    \nonumber \\
\widetilde{6} &\equiv& \{22, 36, 50, 79, 93, 107, 121, 132, 146, 160, 182, 203, 217, 239, 253\}, \nonumber \\
\widetilde{7} &\equiv& \{22, 36, 50, 79, 93, 107, 121, 139, 153, 175, 189, 196, 210, 224, 246\},  \nonumber \\
\widetilde{8} &\equiv& \{15, 22, 29, 36, 43, 50, 57, 64, 79, 86, 93, 100, 107, 114, 121\}, \nonumber \\
\widetilde{9} &\equiv& \{15, 22, 29, 36, 43, 50, 57, 128, 143, 150, 157, 164, 171, 178, 185\}, \nonumber \\
\widetilde{10} &\equiv& \{15, 22, 29, 36, 43, 50, 57, 192, 207, 214, 221, 228, 235, 242, 249\}, \nonumber \\
\widetilde{11} &\equiv& \{22, 36, 50, 68, 82, 96, 118, 139, 153, 175, 189, 207, 221, 235, 249\},  \nonumber \\
\widetilde{12} &\equiv& \{22, 36, 50, 68, 82, 96, 118, 143, 157, 171, 185, 203, 217, 239, 253\}. \nonumber
\end{eqnarray*}
\begin{figure}[pth!]
\centerline{\includegraphics[width=15.cm,clip=]{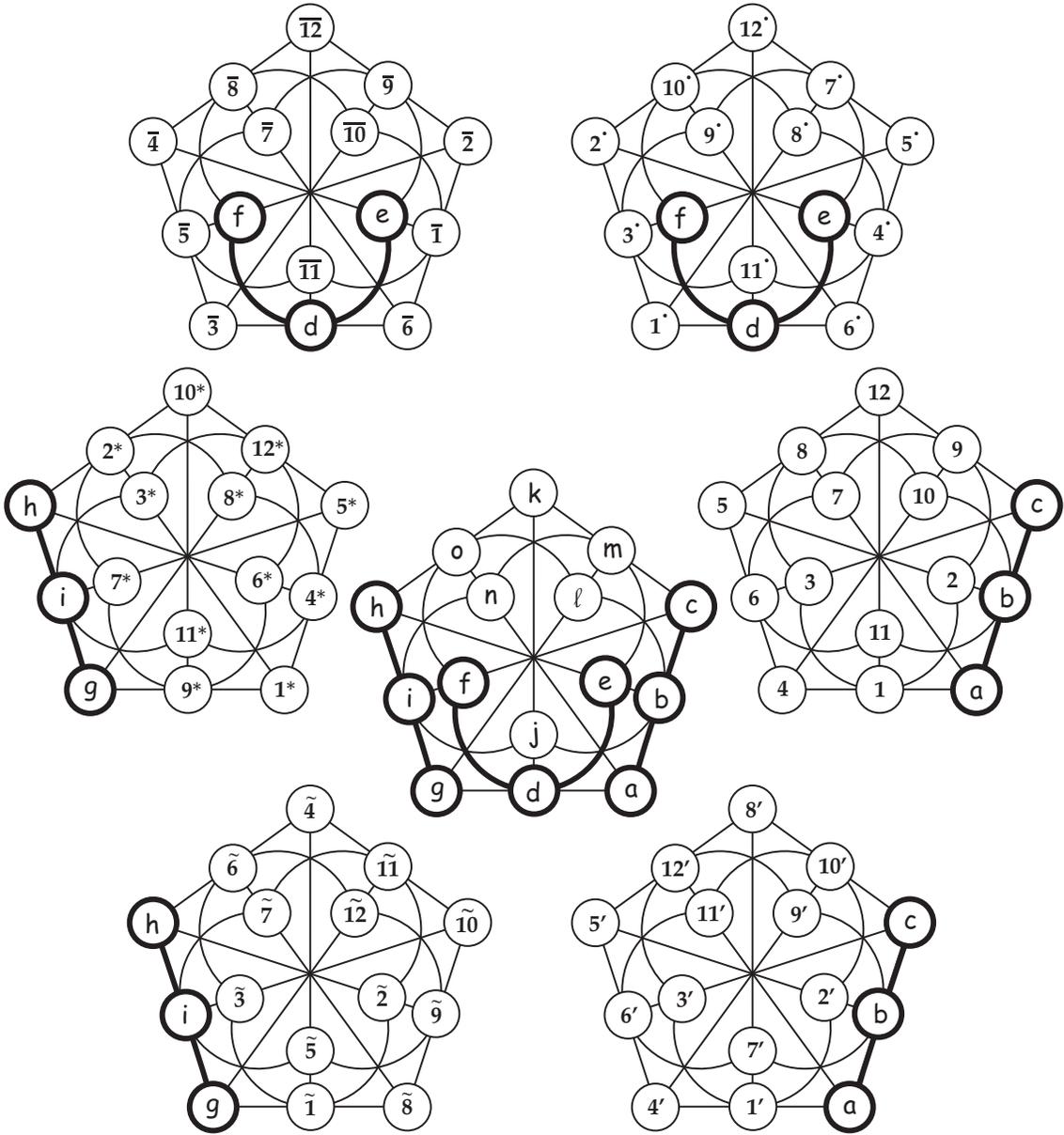}}
\vspace*{.2cm}
\caption{A schematic sketch of the geometry behind the qubit-quoctic generalized Pauli group. The seven doilies lie in three pencils on the distinguished doily (middle); the carrier lines of these pencils are shown in boldface.}
\end{figure}
\begin{figure}[htb]
\centerline{\includegraphics[width=4.cm,clip=]{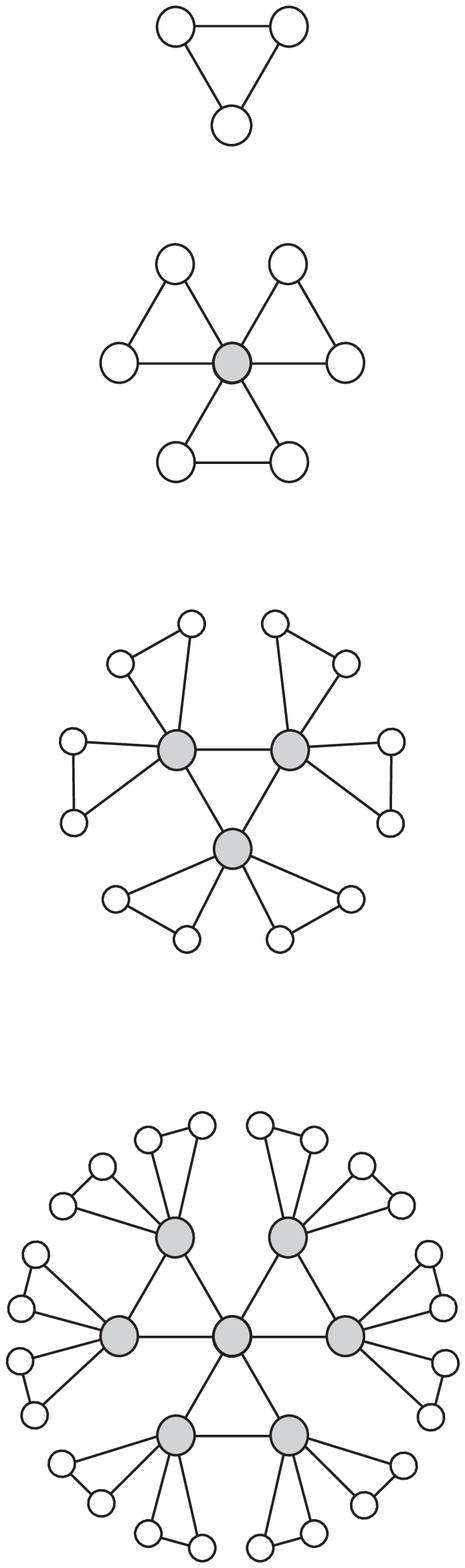}}
\vspace*{.2cm}
\caption{A compact illustration of the properties of the above-described finite geometries for (top to bottom) $k = 2, 3, 4$ and $5$. Each circle represents a doily and each triangle stands for a pencil of doilies; the filled circle means that the corresponding doily contains solely exceptional points.}
\end{figure}
\noindent
One observes that each twelve-element set features three common elements and the two sets in a pair share a single element, this being the element $4 = I_{(2)} \otimes X_{(8)}^4$, $32 = I_{(2)} \otimes Z_{(8)}^4$ and $36 = I_{(2)} \otimes Z_{(8)}^4 X_{(8)}^4$, respectively. Paralleling the preceding case, we again form the point-line incidence geometry where points are the 87 maximum sets, but where two points are now collinear if the corresponding sets have exactly \emph{seven} (=$2^3 - 1$) elements in common. This geometry, depicted in Figure 3, consists  of seven doilies. These form three pencils, each comprising the doily generated by the 15 exceptional sets and two doilies coming from pair $A$ if the missing line $\{a, b, c\}$ is added, from pair $B$ if the missing line $\{d, e, f\}$ is added and from pair $C$ if the missing line $\{g, h, i\}$ is added. The three lines in questions are pairwise skew; for the reader well-acquainted with the structure of the doily it may be interesting to learn  that  in the dual of the distinguished doily these lines answer to a \emph{tricentric} triad of points.

\par
For the next case in the hierarchy, $k = 4$ (qubit-quhexadecit), our computer calculations showed that 1023 elements of the corresponding Pauli group form 183 maximum sets of 31 elements each, out of which 39 are exceptional.
We shall not go into much detail here but mention only the outcome of our analysis, illustrated in line three of Figure 4. The geometry is defined as in the previous two cases, save for the fact that collinearity is now synonymous with sharing \emph{fifteen} (= $2^4 - 1$) elements. The three ``central" doilies in a pencil answer to the 39 elements  of the exceptional set, with the carrier line accounting for the finer, 36+3 split; hence, this subgeometry is isomorphic to the full $k=2$ case (see Figure 4).  The 12 doilies at sides, underlying 144 ordinary points, form three quadruples, each being a ``satellite" to one central doily. A quadruple further splits into two pairs, and the two doilies in each pair form a pencil with the corresponding ``central" doily. So we have altogether seven pencils of doilies.

\par In the $k = 5$ case, we find 4095 elements of the associated Pauli group forming 375 maximum sets of 63 elements each. The 87 of them are exceptional, and their geometry is fully isomorphic to the $k = 3$ one (see Figure 4). The 24 satellite doilies account for the remaining 288 ordinary points. They form 12 pairs and with `six `central" doilies define 12 pencils; taking into account three ``central" pencils, this case is thus seen to be altogether endowed with 15 pencils.

\par
These four cases obviously suffice to infer the geometry for an arbitrary $k \geq 2$. This geometry, which features $2^{k-2}36$ ordinary points, will comprise $2^{k} - 1$ copies of the generalized quadrangle of order two that form $2^{k-1} - 1$ pencils arranged into a remarkable nested configuration. The core of this configuration consists of exceptional points and is isomorphic to the full $k-2$ geometry. From Figure 4 it can readily be discerned how to construct  generically the $k+2$ geometry around the $k$ one. One simply takes a ``peripheral" doily, i.\,e. any doily that does not consist of exceptional points only, and associate with it two pairs of doilies in such manner that they form two pencils. Obviously, we shall get two distinct sequence patterns according as the initial geometry is that of the $k=2$ or $k=3$ case.

\section{Conclusion}
Following and extending the strategy adopted in \cite{pl2}, a finite geometrical treatment of the generalized Pauli groups associated with the Hilbert spaces of type $\bC^2 \otimes \bC^{2^k}$, where an integer $k \geq 2$, has been performed. A point-line incidence structure was defined as follows: its points are maximum sets of pairwise commuting group elements and two distinct points are collinear if the corresponding sets share $2^{k} - 1$ elements. The points were found to be of two distinct kinds, referred to as ordinary and exceptional; a point of the latter kind being represented by such maximum set that shares with any other maximum set at least  one element. A computer-based analysis of the first four cases in the sequence implies that, in general, this  geometry features $2^{k} - 1$ copies of the doily that form $2^{k-1} - 1$ pencils arranged into a remarkable nested configuration reminding a fractal-like behaviour with growing $k$. The core of this geometry, being generated by exceptional points, is isomorphic to the full geometry of the $k-2$ case. Finding the doily, the smallest \emph{triangle-free} $v_3$-configuration, to be the fundamental building block of these geometries is pleasing of its own, all the more that --- as already stressed in introduction --- this generalized quadrangle is the geometry behind the generalized Pauli group of {\it two-qubits} and it plays also an essential role in finite geometrical aspects of the still mysterious black-hole/qubit correspondence. We hope that all the above-discussed properties will soon be given a rigorous, computer-free proof.

\section*{Acknowledgements}
This work was partially supported by the VEGA grant agency projects 2/0092/09 and 2/0098/10. We are extremely grateful to Dr. Petr Pracna for providing us with improved electronic versions of all the figures.

\end{document}